\documentclass{emulateapj}
\usepackage{natbib}
\usepackage{amsmath}

\begin{document}
\title{Systemic Proper Motions of Milky Way Satellites from Stellar Redshifts: the Carina, Fornax, Sculptor and Sextans Dwarf Spheroidals\footnote{This paper includes data gathered with the 6.5-m Magellan Telescopes at Las Campanas Observatory, Chile.}}
\shorttitle{dSph Proper Motions}
\author{Matthew G. Walker\altaffilmark{1,2}, Mario Mateo\altaffilmark{2} and Edward W. Olszewski\altaffilmark{3}}
\email{walker@ast.cam.ac.uk}
\altaffiltext{1}{Institute of Astronomy, University of Cambridge, UK}
\altaffiltext{2}{Department of Astronomy, University of Michigan, Ann Arbor}
\altaffiltext{3}{Steward Observatory, The University of Arizona, Tucson, AZ}

\begin{abstract} 

The transverse motions of nearby dwarf spheroidal (dSph) galaxies contribute line-of-sight components that increase with angular distance from the dSph centers, inducing detectable gradients in stellar redshift.  In the absence of an intrinsic velocity gradient (e.g., due to rotation or streaming), an observed gradient in the heliocentric rest frame (HRF) relates simply to a dSph's systemic proper motion (PM).  Kinematic samples for the Milky Way's brightest dSph satellites are now sufficiently large that we can use stellar redshifts to constrain systemic PMs independently of astrometric data.  Data from our Michigan/MIKE Fiber System (MMFS) Survey reveal significant HRF velocity gradients in Carina, Fornax and Sculptor, and no significant gradient in Sextans.  Assuming there are no intrinsic gradients, the data provide a relatively tight constraint on the PM of Fornax, $(\mu_{\alpha}^{HRF},\mu_{\delta}^{HRF})=(+48 \pm 15,-25\pm 14)$ mas century$^{-1}$, that agrees with published HST astrometric measurements.  Smaller data sets yield weaker constraints in the remaining galaxies, but our Carina measurement, $(\mu_{\alpha}^{HRF},\mu_{\delta}^{HRF})=(+25\pm 36,+16\pm 43)$ mas century$^{-1}$, agrees with the published astrometric value.  The disagreement of our Sculptor measurement, $(\mu_{\alpha}^{HRF},\mu_{\delta}^{HRF})= (-40 \pm 29, -69 \pm 47)$ mas century$^{-1}$, with astrometric measurements is expected if Sculptor has a rotational component as reported by \citet{battaglia08}.  For Sextans, which at present lacks an astrometric measurement, we measure $(\mu_{\alpha}^{HRF},\mu_{\delta}^{HRF})=(-26 \pm 41, +10 \pm 44)$ mas century$^{-1}$.

\end{abstract}
\keywords{galaxies: dwarf ---  (galaxies:) Local Group ---  galaxies: individual (Carina, Fornax, Sculptor, Sextans)}

\section{Introduction}

In recent years there has been an explosion in the quantity of kinematic data available for individual stars in the brightest Local Group dwarf spheroidal (dSph) galaxies (e.g., \citealt{battaglia06,koch07,mateo08,walker08a}).  Many of the major surveys producing such data are designed to quantify the amount and distribution of dark matter in these systems, but the improved statistics benefit other investigations as well.  We now know that what were once thought to be simple dSph stellar populations often contain multiple components separable by age, metallicity and kinematics \citep{tolstoy04,battaglia08}.  Some dSphs display evidence for localized kinematic substructure \citep{kleyna03,walker06b}, while some exhibit tidal streaming motion among their outermost stars \citep{munoz06,sohn07,mateo08}.  Taken together, all these results inform models of galaxy evolution at the smallest scales.  

Here we put the available kinematic data sets to another new use---measuring systemic proper motions (PMs) of dSphs independently of astrometric data.  \citet{kaplinghat08} have recently argued that one can use line-of-sight (LOS) velocity samples of $\ge 1000$ stars to measure PM with precision similar to that which is achieved with HST astrometry, while samples of $\ge 5000$ can do significantly better.  The technique relies on the detection of ``perspective rotation'': at large angular radii, a dSph's transverse motion has a non-negligible component along the LOS, causing spectra to become increasingly redshifted along the direction of the PM vector.  In the absence of an intrinsic velocity gradient (e.g., due to true rotation and/or tidal streaming), the strength and orientation of an observed gradient relates simply to the systemic PM.  This effect is frequently considered in kinematic studies of the Magellanic Clouds (e.g., \citealt{feast61,vandermarel02}), and has recently been used to constrain the PM of M31 from the velocities of its satellites \citep{vandermarel07}.  

Here we apply the perspective technique for the first time to large dSph velocity samples and present the resulting constraints on the PMs of Carina, Fornax, Sculptor and Sextans.  The data result from our Magellan/MMFS observations as of 2008 August and are presented elsewhere \citep[Paper II]{walker08a}.  To our MMFS data we add the Fornax kinematic data of \citet{walker06a}, which contribute 155 stars not observed with MMFS.  We do not combine with other published data, but have verified that, with one exception that owes to different spatial sampling (see Section \ref{sec:discussion} discussion of Carina), doing so would leave our results qualitatively unchanged.  The samples contain velocity measurements for the following numbers of stars: Carina:  1982 stars (774 members); Fornax 2793 stars (2610 members); Sculptor: 1541 stars (1365 members); Sextans: 947 stars (441 members).  

\section{Velocity Gradients}
\label{sec:gradients}

We first demonstrate that the available kinematic data exhibit significant velocity gradients.  For each dSph, Paper II presents LOS stellar velocities, $V$, measured in the heliocentric rest frame (HRF).  Allowing for an HRF velocity gradient $k\equiv dV/dR'$, where $R'$ is angular distance from the dSph center in the direction of the gradient, the data have likelihood
\begin{eqnarray}
  L(\langle V\rangle,\sigma_{V_0},k)\propto \displaystyle\prod_{i=1}^N
  \biggl (\frac{1}{\sqrt{(\sigma_{V_0}^2+\sigma_{V_i}^2)}}\hspace{1in}\\
  \times \exp\biggl [-\frac{1}{2}\frac{(V_i-\langle V \rangle -kR_i\cos(\theta_i-\theta_{0}))^2}{\sigma_{V_0}^2+\sigma_{V_i}^2} \biggr ]\biggr )\nonumber,
  \label{eq:protation}
\end{eqnarray}
where $\sigma_{V_i}$ is the measurement error, $\sigma_{V_0}$ is the internal velocity dispersion, and $\theta_i$, $\theta_{0}$ are position angles of the star and gradient, respectively.  \citet[Paper III hereafter]{walker08b} introduce an algorithm that evaluates the membership probability, $P_{M}$, of each star according to its velocity, magnesium index and position.  Assigning weights to the data points according to their membership probabilities, the expected log-likelihood is
\begin{eqnarray}
  E(\ln L)=-\frac{1}{2}\displaystyle\sum_{i=1}^N P_{M_i}\ln (\sigma_{V_0}^2+\sigma_{V_i}^2)\hspace{1.2in}\\
-\frac{1}{2}\displaystyle\sum_{i=1}^N P_{M_i} \biggl [\frac{(V_i-\langle V \rangle -kR_i\cos(\theta_i-\theta_{0}))^2}{\sigma_{V_0}^2+\sigma_{V_i}^2} \biggr ]
+\mathrm{const}.\nonumber
  \label{eq:explogprotation}
\end{eqnarray}

For a given $\theta_{0}$, the estimates\footnote{$\hat{X}$ denotes the estimate of $X$.} $\langle \hat{V}\rangle$, $\hat{\sigma}_{V_0}$ and $\hat{k}$ take the values that maximize $E(\ln L)$.  Following \citet{walker08b}, we obtain estimates by setting equal to zero the partial derivative of $E(\ln L)$ with respect to each parameter, and then solving iteratively.  For example, $\hat{k}$ is calculated in each iteration as
\begin{equation}
  \hat{k}=\frac{\displaystyle \sum_{i=1}^N\frac{\hat{P}_{M_i}[V_i-\langle \hat{V} \rangle]R_i\cos(\theta_i-\theta_0)}{1+\sigma_{V_i}^2/\hat{\sigma}_{V_0}^2}}{\displaystyle \sum_{i=1}^N\frac{\hat{P}_{M_i}[R_i\cos(\theta_i-\theta_0)]^2}{1+\sigma_{V_i}^2/\hat{\sigma}_{V_0}^2}}.
  \label{eq:k}
\end{equation}

For each dSph we consider possible position angles for the velocity gradient of $\theta_{0}=\{0^{\circ},3^{\circ},6^{\circ},...,180^{\circ}\}$.  We quantify the significance of the maximum gradient, $\hat{k}_{max}$, via Monte Carlo simulation using permutations of the real $(V_i,\sigma_{V_i},R_i,\theta_i)$ data.  In each of $1000$ realizations we re-assign $(V_i,\sigma_{V_i},\hat{P}_{M_i})$ tuples randomly to $(R_i,\theta_i)$ pairs.  Thus the simulated data sets have the same spatial sampling and overall velocity distribution as the real data, but scramble any existing correlation between velocity and position.  We define the significance, $p_{\hat{k}_{max}}$, of the maximum velocity gradient measured from the real data to be the fraction of simulated data sets that fail to produce, at \textit{any} position angle, a gradient as large as the maximum gradient observed in the real data.  

Table \ref{tab:global} lists for each dSph the values of $\hat{k}_{max}$ and $p_{\hat{k}_{max}}$ corresponding to the position angle, $\theta_{0_{max}}$, that gives the largest velocity gradient.  Carina, Fornax and Sculptor all exhibit HRF velocity gradients that are significant at the $p_{\hat{k}_{max}}>0.973$ level.  Sextans shows no significant gradient, with $p_{\hat{k}_{max}}=0.753$.  Despite their significance, none of the gradients inflate estimates of the global velocity dispersion---estimates $\langle \hat{V}\rangle $ and $\hat{\sigma}_{V_0}$ are identical to those obtained in the analysis of Paper III, where we assume no velocity gradient.  

\section{dSph Proper Motions}
\label{sec:propermotion}

The observed velocity gradients can result from the perspective effect and/or some combination of intrinsic rotation and streaming motions.  Here we assume that the latter two phenomena are negligible, in which case a gradient reflects the transverse motion of the dSph.  Suppose $v_{rel}(\alpha,\delta)$ is the projection of the relative motion between Sun and dSph along the line of sight defined by equatorial coordinates $(\alpha,\delta)$.  To an observer instantaneously at the Sun but comoving with the dSph---i.e., in the dSph rest frame (DRF)---a dSph star at $(\alpha,\delta)$ has LOS velocity 
\begin{equation}
  V_{DRF}=V-v_{rel}(\alpha,\delta),
  \label{eq:dsphframe}
\end{equation}
where $V$ is the star's HRF velocity.  In Appendix \ref{app:drfgrf} we derive formulae that express $v_{rel}(\alpha,\delta)$ in terms of the components $(\mu_{\alpha},\mu_{\delta})$ of the dSph's systemic PM.  

We assume a given dSph has a Gaussian velocity distribution with mean $\langle V\rangle_{DRF}=0$ and variance $\sigma_{V_0}^2$, with no intrinsic gradient.  The HRF gradient (Section \ref{sec:gradients}) then reflects the smooth variation in $v_{rel}$ across the face of the dSph, and the likelihood function becomes
\begin{eqnarray}
  L(\mu_{\alpha},\mu_{\delta})\propto 
  \displaystyle\prod_{i=1}^N \biggl (\frac{1}{(\sigma_{V_0}^2+\sigma_{V_i}^2)}\hspace{1.5in}\\
  \times \exp\biggl [-\frac{1}{2}\frac{[V_i-v_{rel}(\alpha_i,\delta_i)]^2}{\sigma_{V_0}^2+\sigma_{V_i}^2}\biggr ]\biggr ).\nonumber
  \label{eq:pmlikelihood}
\end{eqnarray}
Again assigning weights to each star according to probability of dSph membership, we obtain the expected log-likelihood
\begin{eqnarray}
  E(\ln L)= -\frac{1}{2}\displaystyle\sum_{i=1}^N P_{M_i}\ln (\sigma_{V_0}^2+\sigma_{V_i}^2) \hspace{1in}\\
-\frac{1}{2}P_{M_i}\biggl [-\frac{1}{2}\frac{[V_i-v_{rel}(\alpha_i,\delta_i)]^2}{\sigma_{V_0}^2+\sigma_{V_i}^2}\biggr ]
+\mathrm{const}.\nonumber
  \label{eq:pmloglikelihood}
\end{eqnarray}

We measure PM by maximizing $E(\ln L)$ over the parameter pair $(\mu_{\alpha},\mu_{\delta})$, which specifies $v_{rel}(\alpha_i,\delta_i)$ according to Equations \ref{eq:axayaz1} - \ref{eq:vrelagain}.  During this procedure we hold the velocity dispersion $\sigma_{V_0}$ fixed at the value that maximized the expected log-likelihood in Equation \ref{eq:explogprotation}.  These values are $\sigma_{V_0}=6.6$ km s$^{-1}$ (Carina), $11.6$ km s$^{-1}$ (Fornax), $9.2$ km s$^{-1}$ (Sculptor) and $7.9$ km s$^{-1}$ (Sextans).

\begin{figure}
  \epsscale{1.25}
  \plotone{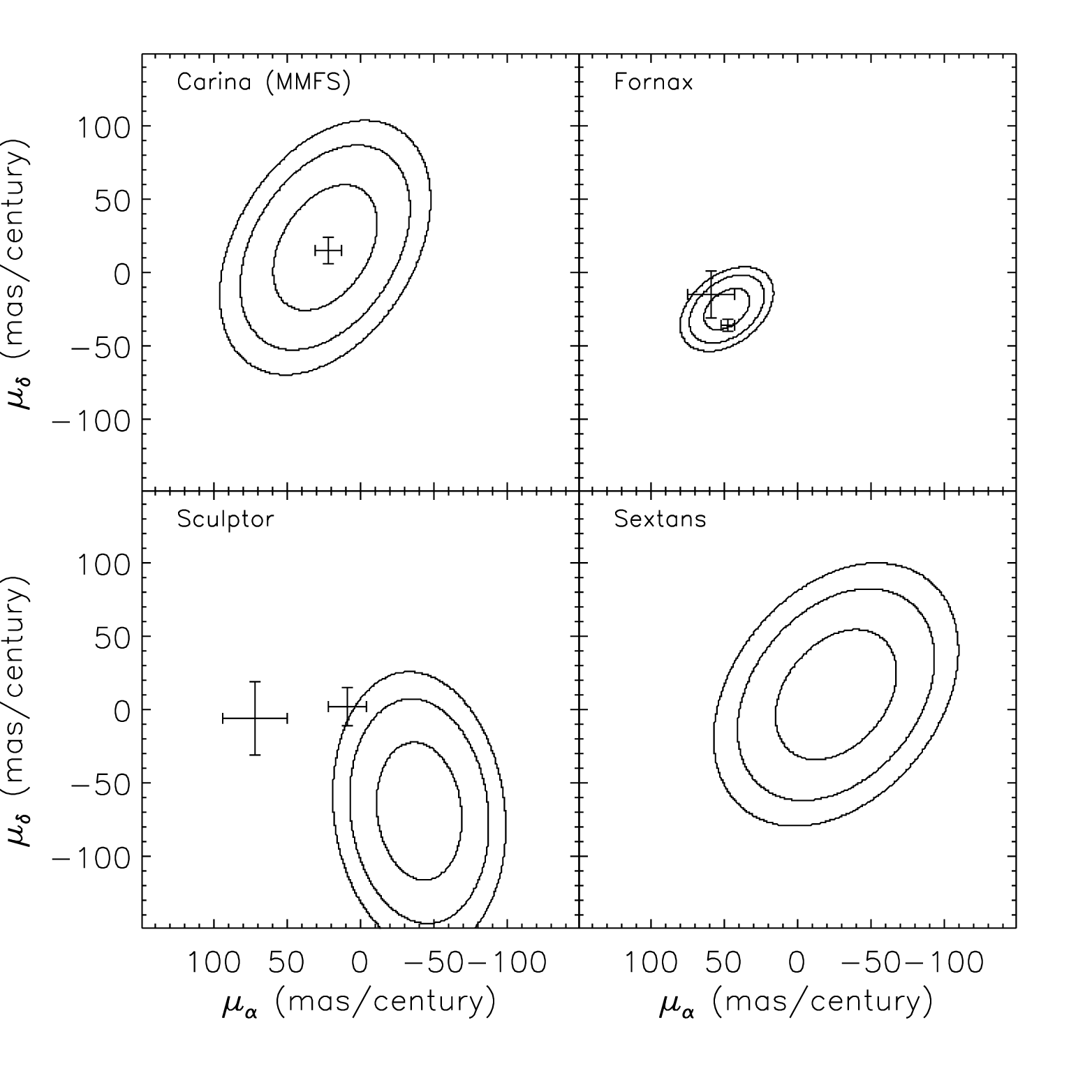}
  \caption{\scriptsize Proper motions measured from stellar redshift data.  Contours indicate $1\sigma$, $2\sigma$ and $3\sigma$ constraints on the PM of each dSph, obtained by evaluating Equation \ref{eq:pmloglikelihood} using the MMFS velocity data.  Points with error bars indicate published PM measurements derived from precise astrometry (Carina---\citealt{piatek03}; Fornax---\citealt{dinescu04,piatek07}; Sculptor---\citealt{schweitzer95,piatek06}).  Where two measurements exist, the HST measurement has smaller error bars.}
  \label{fig:propermotion}
\end{figure}

Figure \ref{fig:propermotion} displays and Table \ref{tab:global} lists the resulting PM measurements.  Contours in the $\mu_{\alpha}^{HRF},\mu_{\delta}^{HRF}$ plane of Figure \ref{fig:propermotion} enclose regions containing $68\%$, $95\%$ and $99\%$, respectively, of the volume underneath the $\exp[E(\ln L)]$ surface.  Points with errorbars indicate published, astrometric PM measurements \citep{schweitzer95,piatek03,dinescu04,piatek06,piatek07}.  

For two (Carina and Fornax) of the three dSphs with astrometric PMs, the PMs derived from MMFS data agree well with astrometric values.  We obtain our tightest constraint on the PM of Fornax, for which we measure $(\mu_{\alpha}^{HRF},\mu_{\delta}^{HRF})= (+48 \pm 15,-25\pm 14)$ mas century$^{-1}$.  The $1\sigma$ error ellipse in Figure \ref{fig:propermotion} overlaps the measurements of both \citet{dinescu04} and \citet{piatek07}.  It should be noted that the Fornax measurement by \citet{piatek07}, based on HST astrometry, has the best precision of any dSph PM measurement ($\pm 5$ mas century$^{-1}$).  It is encouraging that our measurement of Fornax agrees with the \citet{piatek07} value and has precision similar to that of HST measurements for other dSphs.  

Our measurement for Sculptor disagrees with the two published astrometric PMs \citep{schweitzer95,piatek06}, which also disagree with each other.  However, if either of the astrometric PMs is correct, disagreement with our result is to be expected.  If we use Equation \ref{eq:dsphframe} to remove the perspective effect due to either of the astrometric PMs, the resulting Sculptor velocities exhibit a strong residual DRF velocity gradient ($\hat{k}\geq 6.0$ km s$^{-1}$deg$^{-1}$; $\hat{p}_{\hat{k}_{max}}\geq 0.994$).  Thus, if either of the astrometric PMs is correct, our velocity data imply that Sculptor has a strong intrinsic velocity gradient.  Whether attributable to a rotational component, as argued by \citet{battaglia08} on the basis of independent kinematic data, or to tidal streaming, such a gradient would invalidate the assumptions inherent to our PM measurement.  
\renewcommand{\arraystretch}{0.6}
\begin{deluxetable}{lrrrrrrrrrrrrrrrrr}
  \tabletypesize{\scriptsize}
  \tablecaption{ \scriptsize HRF Velocity Gradients and Proper Motions}
  \tablehead{\\
\colhead{dSph}&\colhead{$\hat{k}_{max}$}&\colhead{$\theta_{0,max}$}&\colhead{$p_{\hat{k}_{max}}$}&\colhead{$\mu_{\alpha}$}&\colhead{$\mu_{\delta}$}\\
    \colhead{}&\colhead{(km/s/deg)}&\colhead{(deg)}&\colhead{}&\colhead{(mas/cent)}&\colhead{(mas/cent)}}
  \startdata
\\
Car&$2.5\pm 0.8$&21&0.973&$+25\pm 36$&$+16\pm 43$\\
For&$6.3\pm 0.2$&120&$>0.999$&$+48 \pm 15$&$-25 \pm 14$\\
Scl&$-5.5\pm 0.5$&21&$>0.999$&$-40 \pm 29$&$-69 \pm 47$\\
Sex&$-2.1\pm 0.8$&120&0.753&$-26 \pm 41$&$+10 \pm 44$\\
\enddata
  \label{tab:global}
\end{deluxetable}

\section{Discussion}  
\label{sec:discussion}

We have presented the first constraints on dSph PMs that are derived from stellar redshifts.  Given the sizes of our kinematic samples, we are able to measure PM with approximately the precision predicted from simulations by \citet[K08 hereafter]{kaplinghat08}.  K08 adopt a more sophisticated model for a dSph's intrinsic velocity distribution, marginalizing over several parameters that describe the gravitational potential and orbital anisotropy.  However, our simple assumption that the dSph velocities are drawn everywhere from a single, isotropic Gaussian distribution is consistent with the flat empirical velocity dispersion profiles of dSphs \citep{walker07b}, and yields similar constraints as the K08 method.  

\citet{battaglia08} conclude that Sculptor rotates, based on their detection of a significant LOS velocity gradient in the Milky Way rest frame.  We detect a similar gradient, but note that it can be accounted for entirely by perspective effects, provided that Sculptor's PM differs from both of the astrometric measurements.  In order to examine the implications of our PM measurement for a non-rotating Sculptor's orbit, we have integrated the orbit backward in time for 6 Gyr using code provided by Slawomir Piatek.  Assuming a static MW potential with disk, spheroid and halo components identical to those adopted by \citet{johnston99}, our measurement implies a plausible Sculptor orbit ($r_{peri}=66_{-45}^{+12}$ kpc, $r_{apo}=149_{-63}^{+97}$ kpc) with eccentricity similar to that implied by the HST astrometric PM ($r_{peri}=68_{-37}^{+15}$,$r_{apo}=122_{-25}^{+91}$ kpc; \citealt{piatek06}).  

Conclusions regarding Sculptor's status then depend on the initial assumption.  If one assumes and corrects for (either of) the published astrometric PMs, the significant residual DRF velocity gradient implies a strong intrinsic gradient that is likely due to some combination of rotation and/or tidal streaming (see \citealt{battaglia08}).  If one assumes instead that Sculptor has no intrinsic velocity gradient, then the HRF gradient implies a PM that is inconsistent with either of the published astrometric measurements.

\citet{munoz06} detect a strong LOS velocity gradient from Carina members projected at large angular distances ($R> 1^{\circ}$), well outside the region sampled by MMFS.  Had we combined their extended kinematic sample with our MMFS data, we would have measured a PM of $(\mu_{\alpha}^{HRF},\mu_{\delta}^{HRF})=(+56\pm 29,-1\pm 33)$, in poorer agreement with astrometric measurements.  \citet{munoz06} argue convincingly that the outermost regions of Carina exhibit tidal streaming motions, which would invalidate our PM measurement that includes their outer sample.  It is encouraging that in the inner regions where no evidence for tides exists, our PM measurement stands in excellent agreement with the astrometric measurement.  In any case, Carina and Sculptor help illustrate that PMs measured from stellar redshifts ought to be regarded with caution, as they are only as valid as the assumption of no intrinsic gradient.  The agreement, to high precision, of our Fornax PM with the astrometric measurement serves as strong evidence that Fornax has no intrinsic velocity gradient over the kinematically sampled region.

Our results include the first PM measurement of any kind that has been made for Sextans.  Our measurement implies that Sextans, at a distance of 86 kpc \citep{mateo98}, is receding from its perigalacticon distance of $r_{peri}=66_{-61}^{+17}$ kpc toward an apogalactic distance of $r_{apo}=129_{-33}^{+113}$.  We note, however, that the relatively large error bars on Sextans' PM accommodate orbital eccentricities ranging from $0.25- 0.89$ within the $95\%$ confidence interval.  The most radial of these orbits would bring Sextans within $\sim 5$ kpc of the Galactic center.  If one takes the most likely PM at face value, it becomes unlikely that Sextans is a member of a stream associated with other well-known objects in the Milky Way Halo.  None of the PMs predicted by \citet{lyndenbell95} for possible Sextans associations with Sculptor, Sculptor/Fornax, or Pal 3/Pal 2/NGC 5824 falls within even the large error bars of the measured PM (although none of the stream associations is ruled out beyond the $2\sigma$ level).

We thank Louie Strigari for helpful discussion, and Slawomir Piatek and Tad Pryor for sharing the code used to calculate orbital parameters.  EO acknowledges NSF grants AST-0205790, 0505711, and 0807498.  MM acknowledges NSF grants 0206081, 0507453, and 0808043.  MGW acknowledges support from the STFC-funded Galaxy Formation and Evolution programme at the Institute of Astronomy, University of Cambridge.  

\appendix
\section{Dwarf Rest Frame (DRF)}
\label{app:drfgrf}

Suppose $\mathbf{A}_{D}$ and $\mathbf{A}_{*}$ are the (three dimensional) position vectors of a dSph and one of its stars, respectively, specified in a coordinate system with origin at the Sun.  The relative motion between Sun and dSph along the line of sight to a star whose equatorial coordinates are $(\alpha_*,\delta_*)$ is the scalar projection
\begin{equation}
  v_{rel}(\alpha_*,\delta_*)=\frac{\mathbf{A}_*}{A_*}\cdot(\mathbf{\dot{A}}_{D}),
  \label{eq:vrel}
\end{equation}
where $\mathbf{\dot{A}} \equiv d\mathbf{A}/dt$.

In order to evaluate Equation \ref{eq:vrel} we define Cartesian coordinates such that the $+x$, $-y$, and $+z$ axes point toward $(\alpha,\delta)=(6^{h},0^{\deg})$, toward the vernal equinox, and toward the North Celestial Pole, respectively.  Thus the position of a star is fully specified by $\mathbf{A}_*=(A_{*_x}\hat{x}+A_{*_y}\hat{y}+A_{*_z}\hat{z})$, and the unit vector $\mathbf{B}_* \equiv \mathbf{A}_*/A_*$ has components
\begin{eqnarray}
  B_{*_x}=\cos(\delta_*)\sin(\alpha_*);\\
  \label{eq:axayaz1}
  B_{*_y}=-\cos(\delta_*)\cos(\alpha_*);\\
  B_{*_z}=\sin(\delta_*).
\end{eqnarray}
The (three dimensional) velocity of a dSph at distance $A_{D}$, with HRF line-of-sight velocity $V_{D}=\dot{A}_{D}$ and HRF proper motion $(\mu_{\alpha},\mu_{\delta})$ has components
\begin{eqnarray}
  \dot{A}_{D_x}=V_{D}\cos(\delta_{D})\sin(\alpha_{D})
  +A_{D}\mu_{\alpha}\cos(\delta_{D})\cos(\alpha_{D})
  -A_{D}\mu_{\delta}\sin(\delta_{D})\sin(\alpha_{D});\\
  \dot{A}_{D_y}=-V_{D}\cos(\delta_{D})\cos(\alpha_{D})
  +A_{D}\mu_{\delta}\sin(\delta_{D})\cos(\alpha_{D})
  +A_{D}\mu_{\alpha}\cos(\delta_{D})\sin(\alpha_{D});\\
  \dot{A}_{D_z}=V_{D}\sin(\delta_{D})+A_{D}\mu_{\delta}\cos(\delta_{D}).
  \label{eq:vxvyvz}
\end{eqnarray}
Thus the relative motion between dSph and Sun projects along the star's line of sight with magnitude
\begin{equation}
  v_{rel}(\alpha_*,\delta_*)=\mathbf{B}_*\cdot\mathbf{\dot{A}}_{D}=B_{*_x}\dot{A}_{D_x}+B_{*_y}\dot{A}_{D_y}+B_{*_z}\dot{A}_{D_z}.
  \label{eq:vrelagain}
\end{equation}

\bibliography{ref}
\end{document}